\documentclass[intlimits,twoside,a4paper]{article}

\usepackage[cp1251]{inputenc}

\usepackage[eqsecnum]{cmpj3}

\issue{2024}{27}{4}{43601}
\doinumber{10.5488/CMP.27.43601}

\title[Critical, compensation and hysteresis behaviors studies in the ferrimagnetic Blume--Capel model]%
{Critical, compensation and hysteresis behaviors studies in the ferrimagnetic Blume--Capel model with mixed half-integer spin-(3/2, 7/2): Exact recursion relations calculations}%

\author[M. Kake et al.]{M. Kake\refaddr{label1}, S. I. V. Hontinfinde\refaddr{label2, label3,label5}, M. Karimou\orcid{0000-0002-1422-6479}\refaddr{label1,label4,label5}\thanks{Corresponding author: \email{mounirou.karimou@yahoo.fr}.},  R. Houenou\refaddr{label1}, E. Albayrak\refaddr{label6}, R.~A.~A.~Yessoufou\refaddr{label1,label3}\thanks{Corresponding author: \email{yesradca@yahoo.fr}.}, A. Kpadonou\refaddr{label1,label7}
 }
\addresses{
\addr{label1} Institute of Mathematic and Physical Sciences (IMSP), Dangbo, Benin
\addr{label2}Ecole Nationale Sup\'{e}rieure de G\'{e}nie Math\'{e}matique et Mod\'{e}lisation (ENSGMM) d'Abomey, B\'{e}nin
\addr{label3} University of Abomey-Calavi, Department of Physics, Benin
\addr{label4} Laboratoire des Sciences d'Ing\'{e}nierie et de Math\'{e}matique Appliqu\'{e}e (LSIMA)
\addr{label5} Ecole Nationale Sup\'{e}rieure de G\'{e}nie Energ\'{e}tique et Proc\'{e}d\'{e}s d’Abomey, Benin
\addr{label6} Erciyes
 University, Department of Physics, 38039, Kayseri, Turkey
 \addr{label7}ENS and Laboratory of Physics and Applications (LPA) of Abomey, Benin
}


\date{Received June 07, 2024, in final form July 22, 2024}

\begin{document}
\maketitle
\begin{abstract}
The exact recursion relations are used to study the mixed half-integer spin-(3/2, 7/2) Blume--Capel Ising ferrimagnetic system on the Bethe lattice. Ground-state phase diagrams are computed in the $({D_{A}}/{q|J|}, {D_{B}}/{q|J|})$ plane to reveal different possible ground states of the model.  Using the thermal changes of the order-parameters, interesting temperature dependent phase diagrams are constructed in the ($D_{A}/|J|$, $kT/|J|$), ($D_{B}/|J|$, $kT/|J|$) planes as well as in the ($D/|J|$, $kT/|J|$) plane where $D=D_A=D_B$. It is revealed that the system exhibits first- and second-order phase transitions and compensation temperatures for specific model parameter values. Under the constraint of an external magnetic field, the model also produces multi-hysteresis behaviors as single, double and triple hysteresis cycles. Particularly, the impacts of the ferrimagnetic coupling $J$ on the remanent magnetization and the coercitive fields for selected values of the other physical parameters of the system are pointed out. Our numerical results are qualitatively consistent with those reported in the literature.

\printkeywords
%
\end{abstract}

\section{Introduction}
A significant technological and industrial revolution that humanity has experienced in recent decades is due to the numerous experimental research \cite{ra1, ra2, ra3, ra4} and theoretical studies \cite{ra5, ra6, ra7, ra8} carried out by researchers in fundamental sciences. In condensed matter physics, most of these research has focused on studying the thermodynamic and magnetic properties of various types of materials used in many applications today~\cite{ra9, ra10, ra11}.
Systems composed of mixed half-integer spins~\cite{ra5, ra12, ra13, ra14, ra15, ra16, ra17, ra18, ra19, ra20} are of great importance in condensed matter physics due to the various interesting magnetic properties they exhibit, including critical behaviors, compensation properties, reentrance, and hysteresis. These properties are sought after in these types of systems through various methods such as Monte Carlo simulation \cite{ra12, ra17, ra19, ra20, ra21}, effective field theory \cite{ra6, ra24}, mean-field approximation \cite{ra22}, renormalization group method \cite{ra25}, exact recursion relations \cite{ra23} and so on.

The study of ferrimagnetics systems attracts an increasing attention due to their complex magnetic properties, which are exploited in nanotechnology applications \cite{ra26, ra27, ra28}, especially in data storage devices~\cite{ra29, ra30, ra31, ra32, ra33}. The case of the ferrimagnetic Blume--Capel model with mixed spin-(3/2, 7/2), which has been the subject of several recent investigations, and has shown rich magnetic properties both in the presence and absence of an external magnetic field. The magnetic behavior of this system has been studied using the Monte Carlo simulation method \cite{ra21, ra22}. The bilinear antiferromagnetic interactions between nearest neighbors and ferromagnetic interactions between second neighbors were considered, revealing the presence of multiple hysteresis loops,  exchange bias phenomenon, compensations, and magnetization discontinuities in the system. Other studies have been conducted on the same model without considering the interactions between second neighbors. For instance, the ferrimagnetic complex [Cr(CN)$_{4}$(\textmu-CN)$_{2}$Gd(H$_{2}$O)$_{4}$-(bpy)]$_{\rm n}$.4nH$_{2}$O.1.5nbpy was examined using the mean-field approximation with the Bogoliubov inequality \cite{ra34}, showing compensation behaviors at low temperatures. The effects of an external magnetic field on the total magnetization of this complex were also investigated using Monte Carlo simulations \cite{ra20}, revealing single, double, triple and quintuple loops for different values of the crystal field strengths in the sublattices.

While it is true that the Blume--Capel model with mixed spin-(3/2, 7/2) on a square lattice, with only ferrimagnetic interactions between nearest neighbors, has already been sufficiently studied, yielding interesting results, it remains important to verify and expand upon these results using other calculation methods. Additionally, the study of phase diagrams in the planes formed by temperature and two ionic anisotropies, as well as the effect of ferrimagnetic coupling on coercive field and remanent, which are addressed in this work using the exact recursion relation calculations, have not yet appeared in the literature for our model.

This article is structured as follows: section~\ref{sec-II} is devoted to the description of the model and to the methodology used. Results are presented and discussed in detail in section~\ref{sec-III}, followed by the conclusion in section~\ref{sec-IV}.

\section{Model and formalism}
\label{sec-II}

In this work, we have considered the  Blume--Capel model of ferrimagnetic Ising ($J < 0$) type with mixed spins on the Bethe lattice consisting of a bipartite configuration of sublattices A and B of spins. Each site of sublattice A is occupied by a spin-$3/2$ (spin-$s$) with four discrete values $\pm3/2$ and $\pm1/2$, and  site of sublattice B is occupied by a spin-$7/2$ (spin-$\sigma$) with eight discrete values $\pm7/2$, $\pm5/2$, $\pm3/2$, and $\pm1/2$. The spins of the two sublattices alternate on the lattice, with a spin-$s$ occupying the center. The Hamiltonian describing this system may be written as follows:
\begin{eqnarray}
\label{delta-def1}
H=-J\sum_{<i,j>}s_{i}\sigma_{j}-D_{A}\sum_{i}s_{i}^{2}-D_{B}\sum_{j}\sigma_{j}^{2}-h\bigg(\sum_{i}s_{i}+\sum_{j}\sigma_{j}\bigg),
\end{eqnarray}
where $J$ represents the bilinear interaction parameter between spins of the two sublattices. $D_{A}$ and $D_{B}$ are the crystal fields acting on the sites of sublattices A and B, respectively, and $h$ is the external magnetic field uniformly applied to the lattice.

The approach based on exact recursion relations (ERRs) used in this work is elaborately described in reference~\cite{ra23}. In this approach, the partition function is defined as follows:
\begin{eqnarray}
\label{delta-def2}
Z&=&\sum_{\rm Conf} \re^{-\beta H}= \sum_{\rm Spc}P({\rm Spc})\nonumber\\
&=&\sum_{\{s,\sigma\}}\exp\bigg\{\beta\bigg[J\sum_{<i,j>}s_{i}\sigma_{j}+D_{A}\sum_{i}s_{i}^{2} 
+ D_{B}\sum_{j}\sigma_{j}^{2}
+h\bigg(\sum_{i}s_{i}+\sum_{j}\sigma_{j}\bigg)\bigg]\bigg\},
\end{eqnarray}
where $P({\rm Spc})$ is considered to be an unnormalized probability distribution over spin configurations, ${\rm Spc} \equiv {\sigma, s}$, etc. The probability distribution of spin configurations relative to a central spin $s_{0}$ is expressed as follows:
\begin{eqnarray}
\label{delta-def3}
P(\{s_{0}\})=\exp\big[\beta\big(D_{A}s_{0}^{2}+hs_{0}\big)\big]\prod_{k=1}^{q}Q_{n}\big(s_{0}|\sigma_{1}^{(k)}\big).
\end{eqnarray}
The function $Q_{n}$ in equation~(\ref{delta-def3}) accounts for the  interactions between the central spin and the $q$ nearest neighbor spins $\sigma_{1}^{k}$, forming the $k$-th generation of spins, and is calculated as follows:
\begin{eqnarray}
\label{delta-def4}
Q_{n}\big(s_{0}|\sigma_{1}^{(k)}\big)=\exp\big[\beta\big(Js_{0}\sigma_{1}^{(k)}+D_{B}\big(\sigma_{1}^{(k)}\big)^{2}+h\sigma_{1}^{(k)}\big)\big]\prod_{l=1}^{p}Q_{n-1}\big(\sigma_{1}^{(k)}|s_{2}^{(l)}\big),
\end{eqnarray}
with
\begin{eqnarray}
\label{delta-def5}
Q_{n-1}\big(\sigma_{1}^{(k)}|s_{2}^{(l)}\big)=\exp\big[\beta\big(Js_{2}^{(l)}\sigma_{1}+D_{A}\big(s_{2}^{(l)}\big)^{2}+hs_{2}^{(l)}\big)\big]\prod_{m=1}^{p}Q_{n-2}\big(s_{2}^{(l)}|\sigma_{3}^{(m)}\big)
\end{eqnarray}
and $p=q-1$.

Let us now introduce the partition function per branch of the sublattice $A$ as follows:
\begin{eqnarray}
\label{delta-def6}
g_{n}(\{s_{0}\})=\sum_{\{\sigma_{1}\}}Q_{n}(s_{0}|\{\sigma_{1}\}).
\end{eqnarray}
 By using equation~(\ref{delta-def4}), equation~(\ref{delta-def6}) explicitly becomes:
\begin{eqnarray*}
g_{n}(s_{0})=\sum_{\sigma_{1}}\exp\big[\beta\big(Js_{0}\sigma_{1}+D_{B}\sigma_{1}^{2}+h\sigma_{1}\big)\big]g_{n-1}^{p}(\sigma_{1}).
\end{eqnarray*}
%
%
{Taking into account the different values of $\sigma_{1}$ ($\pm 7/2$, $\pm 5/2$, $\pm3 /2$,$\pm  1/2$), the partial partition function $g_{n}(s_{0})$ explicitly becomes:
\begin{eqnarray}
	\label{delta-def7}
	g_{n}(s_{0})	&=&\exp\big[\beta\big(7Js_{0}/2+49D_{B}/4+7h/2\big)\big]g_{n-1}^{p}(7/2)\nonumber\\ \nonumber
	&+&\exp\big[\beta\big(5s_{0}J/2+25D_{B}/4+5h/2\big)\big]g_{n-1}^{p}(5/2)\\  \nonumber
	&+&\exp\big[\beta\big(3Js_{0}/2+9D_{B}/4+3h/2\big)\big]g_{n-1}^{p}(3/2)\\ \nonumber
	&+&\exp\big[\beta\big(Js_{0}/2+D_{B}/4+h/2\big)\big]g_{n-1}^{p}(1/2)\\ \nonumber	
    &=&\exp\big[\beta\big(-7Js_{0}/2+49D_{B}/4+7h/2\big)\big]g_{n-1}^{p}(7/2)\\ \nonumber
    &+&\exp\big[\beta\big(-5s_{0}J/2+25D_{B}/4+5h/2\big)\big]g_{n-1}^{p}(5/2)\\  \nonumber
    &+&\exp\big[\beta\big(-3Js_{0}/2+9D_{B}/4+3h/2\big)\big]g_{n-1}^{p}(3/2)\\
    &+&\exp\big[\beta\big(-Js_{0}/2+D_{B}/4+h/2\big)\big]g_{n-1}^{p}(1/2). 
\end{eqnarray}
For each of the four values of $s_{0}$ ($\pm3/2$,$\pm1/2$), we can calculate a partial partition function. Thus, for $ s_{0}=\pm3/2 $
\begin{eqnarray}
\label{delta-def8}
g_{n}(\pm3/2)&=&\exp\big[\beta\big(\pm21J/4+49D_{B}/4+7h/2\big)\big]g_{n-1}^{p}(7/2)\nonumber\\ \nonumber
&+&\exp\big[\beta\big(\pm15J/4+25D_{B}/4+5h/2\big)\big]g_{n-1}^{p}(5/2)\\  \nonumber
&+&\exp\big[\beta\big(\pm9J/4+9D_{B}/4+3h/2\big)\big]g_{n-1}^{p}(3/2)\\ \nonumber
&+&\exp\big[\beta\big(\pm3J/4+D_{B}/4+h/2\big)\big]g_{n-1}^{p}(1/2)\\ \nonumber
&+&\exp\big[\beta\big(\mp21J/4+49D_{B}/4-7h/2\big)\big]g_{n-1}^{p}(-7/2)\\ \nonumber
&+&\exp\big[\beta\big(\mp15J/4+25D_{B}/4-5h/2\big)\big]g_{n-1}^{p}(-5/2)\\ \nonumber
&+&\exp\big[\beta\big(\mp9J/4+9D_{B}/4-3h/2\big)\big]g_{n-1}^{p}(-3/2)\\ 
&+&\exp\big[\beta\big(\mp3J/4+D_{B}/4-h/2\big)\big]g_{n-1}^{p}(-1/2).
\end{eqnarray}
 for $ s_{0}=\pm1/2 $
\begin{eqnarray}
\label{delta-def9}
g_{n}(\pm1/2)&=&\exp\big[\beta\big(\pm7J/4+49D_{B}/4+7h/2\big)\big]g_{n-1}^{p}(7/2)\nonumber\\ \nonumber
&+&\exp\big[\beta\big(\pm5J/4+25D_{B}/4+5h/2\big)\big]g_{n-1}^{p}(5/2)\\ \nonumber &+&\exp\big[\beta\big(\pm3J/4+9D_{B}/4+3h/2\big)\big]g_{n-1}^{p}(3/2)\\ \nonumber
&+&\exp\big[\beta\big(\pm J/4+D_{B}/4+h/2\big)\big]g_{n-1}^{p}(1/2)\\ \nonumber
&+&\exp\big[\beta\big(\mp7J/4+49D_{B}/4-7h/2\big)\big]g_{n-1}^{p}(-7/2)\\\nonumber
&+&\exp\big[\beta\big(\mp5J/4+25D_{B}/4-5h/2\big)\big]g_{n-1}^{p}(-5/2)\\ \nonumber
&+&\exp\big[\beta\big(\mp3J/4+9D_{B}/4-3h/2\big)\big]g_{n-1}^{p}(-3/2)\\ 
&+&\exp\big[\beta\big(\mp J/4+D_{B}/4-h/2\big)\big]g_{n-1}^{p}(-1/2).
\end{eqnarray}
By analogy to~(\ref{delta-def6}), one can introduce the partition function per branch of the sublattice $B$ as follows:
\begin{eqnarray*}
g_{n-1}(\sigma_{1})=\sum_{s_{2}}\exp\big[\beta\big(Js_{2}\sigma_{1}+D_{A}s_{2}^{2}+hs_{2}\big)\big]g_{n-2}^{p}(s_{2}).
\end{eqnarray*}
By summing over the four (04) values of $ s_{2} $, one can explicitly get
\begin{eqnarray}
	\label{delta-def10}
	g_{n-1}(\sigma_{1})&=&\exp\big[\beta\big(3J\sigma_{1}/2+9D_{A}/4+3h/2\big)\big]g_{n-2}^{p}(3/2)\nonumber\\ \nonumber
	&+&\exp\big[\beta\big(J\sigma_{1}/2+D_{A}/4+h/2\big)\big]g_{n-2}^{p}(1/2)\\ \nonumber
	&
	+&\exp\big[\beta\big(-3J\sigma_{1}/2+9D_{A}/4-3h/2\big)\big]g_{n-2}^{p}
	(-3/2)\\ 
	&+&\exp\big[\beta\big(-J\sigma_{1}/2+D_{A}/4-h/2\big)\big]g_{n-2}^{p}(-1/2).
\end{eqnarray}
Then, it follows that:
\begin{eqnarray}
\label{delta-def11}
g_{n-1}(\pm7/2)& =&\exp\big[\beta\big(\pm21J/4+9D_{A}/4+3h/2\big)\big]g_{n-2}^{p}(3/2)\nonumber\\ \nonumber
&+&\exp\big[\beta\big(\pm7J/4+D_{A}/4+h/2\big)\big]g_{n-2}^{p}(1/2)\\ \nonumber
&
+&\exp\big[\beta\big(\mp21J/4+9D_{A}/4-3h/2\big)\big]g_{n-2}^{p}
(-3/2)\\ 
&+&\exp\big[\beta\big(\mp7J/4+D_{A}/4-h/2\big)\big]g_{n-2}^{p}(-1/2),
\end{eqnarray}
\begin{eqnarray}
\label{delta-def12}
g_{n-1}(\pm5/2)&=&\exp\big[\beta\big(\pm15J/4+9D_{A}/4+3h/2\big)\big]g_{n-2}^{p}(3/2)\nonumber\\\nonumber &+&\exp\big[\beta\big(\pm5J/4+D_{A}/4+h/2\big)\big]g_{n-2}^{p}(1/2)\\ \nonumber
&
+&\exp\big[\beta\big(\mp15J/4+9D_{A}/4-3h/2\big)\big]g_{n-2}^{p}(-3/2)\\ 
&+&\exp\big[\beta\big(\mp5J/4+D_{A}/4-h/2\big)\big]g_{n-2}^{p}(-1/2),
\end{eqnarray}
\begin{eqnarray}
\label{delta-def13}
g_{n-1}(\pm3/2)&=&\exp\big[\beta\big(\pm9J/4+9D_{A}/4+3h/2\big)\big]g_{n-2}^{p}(3/2)\nonumber\\ \nonumber
&+&\exp\big[\beta\big(\pm3J/4+D_{A}/4+h/2\big)\big]g_{n-2}^{p}(1/2)\\ \nonumber
&+&\exp\big[\beta\big(\mp9J/4+9D_{A}/4-3h/2\big)\big]g_{n-2}^{p}(-3/2)\\ 
&+&\exp\big[\beta\big(\mp3J/4+D_{A}/4-h/2\big)\big]g_{n-2}^{p}(-1/2),
\end{eqnarray}
and
\begin{eqnarray}
\label{delta-def14}
g_{n-1}(\pm1/2)&=&\exp\big[\beta\big(\pm 3J/4+9D_{A}/4+3h/2\big)\big]g_{n-2}^{p}(3/2)\nonumber\\\nonumber
&+&\exp\big[\beta\big(\pm J/4+D_{A}/4+h/2\big)\big]g_{n-2}^{p}(1/2)\\ \nonumber
&+&\exp\big[\beta\big(\mp3J/4+9D_{A}/4-3h/2\big)\big]g_{n-2}^{p}(-3/2)\\ 
&+&\exp\big[\beta\big(\mp J/4+D_{A}/4-h/2\big)\big]g_{n-2}^{p}(-1/2).
\label{eq14}
\end{eqnarray}
}
By utilizing equations~(\ref{delta-def8})--(\ref{delta-def9}) and then equations~(\ref{delta-def11})--(\ref{delta-def14}), one can compute the exact recursion relations, defined, on the one hand, as the ratio of the functions $ g_{n} $ for spin-$3/2$
\begin{eqnarray}
\label{delta-def15}
X_{1}=\frac{g_{n}(3/2)}{g_{n}(-1/2)},\quad X_{2}=\frac{g_{n}(1/2)}{g_{n}(-1/2)},\quad X_{3}=\frac{g_{n}(-3/2)}{g_{n}(-1/2)},
\end{eqnarray}
and, on the other hand, as the ratio of the functions $ g_{n-1} $ for spin-$7/2$
\begin{eqnarray}
\label{delta-def16}
Y_{1}=\frac{g_{n-1}(7/2)}{g_{n-1}(-1/2)},\quad
Y_{2}=\frac{g_{n-1}(5/2)}{g_{n-1}(-1/2)},\quad
Y_{3}=\frac{g_{n-1}(3/2)}{g_{n-1}(-1/2)},
\end{eqnarray}
\begin{eqnarray*}
	Y_{4}=\frac{g_{n-1}(1/2)}{g_{n-1}(-1/2)},\quad
	Y_{5}=\frac{g_{n-1}(-7/2)}{g_{n-1}(-1/2)},\quad
	Y_{6}=\frac{g_{n-1}(-5/2)}{g_{n-1}(-1/2)},\quad
	Y_{7}=\frac{g_{n-1}(-3/2)}{g_{n-1}(-1/2)}.
\end{eqnarray*}
The pursued objective is to obtain the order parameters in terms of these relations in  equations~(\ref{delta-def15}) and~(\ref{delta-def16}). Thus, the magnetization of sublattice A, being considered as the central spin, is given by
\begin{eqnarray}
\label{delta-def17}
M_{A}=Z^{-1}\sum_{\{s_{0}\}}s_{0}P(\{s_{0}\})= Z^{-1}\sum_{s_{0}}s_{0}\exp\big[\beta\big(D_{A}s_{0}^{2}+hs_{0}\big)\big]g_{n}^{q}(s_{0})
\end{eqnarray}
and is derived in terms of exact recursion relations as follows:
\begin{eqnarray}
\label{delta-def18}
M_{A}&=&\big[3\re^{\beta(9D_{A}/4+3h/2)}X_{1}^{q}+\re^{\beta(D_{A}/4+h/2)}X_{2}^{q}\nonumber\\ \nonumber
&-&3\re^{\beta(9D_{A}/4-3h/2)}X_{3}^{q}
- \re^{\beta(D_{A}/4-h/2)}\big]/
\big[2\big(\re^{\beta(9D_{A}/4+3h/2)}X_{1}^{q}\\ 
&+& \re^{\beta(D_{A}/4+h/2)}X_{2}^{q}
+\re^{\beta(9D_{A}/4-3h/2)}X_{3}^{q}
+  \re^{\beta(D_{A}/4-h/2)}\big)\big].
\end{eqnarray}
Similarly, the magnetization of sublattice B is explicitly calculated as follows:
\begin{eqnarray}
\label{delta-def19}
M_{B}&=&\big[7\re^{\beta(49D_{B}/4+7h/2)}Y_{1}^{q}+5\re^{\beta(25D_{B}/4+5h/2)}Y_{2}^{q}+3\re^{\beta(9D_{B}/4+3h/2)}Y_{3}^{q}\nonumber\\  \nonumber
&+&\re^{\beta(D_{B}/4+h/2)}Y_{4}
-7\re^{\beta(49D_{B}/4-7h/2)}Y_{5}^{q}-5\re^{\beta(25D_{B}/4-5h/2)}Y_{6}^{q}
\\ \nonumber
&-&3\re^{\beta(9D_{B}/4-3h/2)}Y_{7}^{q}
-\re^{\beta(D_{B}/4-h/2)}\big]/
\big[2\big(\re^{\beta(49D_{B}/4+7h/2)}Y_{1}^{q}\\ \nonumber
&+&\re^{\beta(25D_{B}/4+5h/2)}Y_{2}^{q}+\re^{\beta(9D_{B}/4+3h/2)}Y_{3}^{q}
+\re^{\beta(D_{B}/4+h/2)}Y_{4}\\ 
&+& \re^{\beta(49D_{B}/4-7h/2)}Y_{5}^{q}
+\re^{\beta(25D_{B}/4-5h/2)}Y_{6}^{q}+\re^{\beta(9D_{B}/4-3h/2)}Y_{7}^{q}
+\re^{\beta(D_{B}/4-h/2)}\big)\big].
\end{eqnarray}
With the expressions of magnetizations established, one can readily determine the equations for the critical temperatures. It is worth noting that a second-order transition occurs when magnetizations vanish continuously, while a first-order transition is characterized by a discontinuity in the nullification of the magnetizations.

At the critical temperatures, the following relations must be satisfied as follows:
\begin{eqnarray}
\label{delta-def20}
M_{A}=3\re^{9\beta_{c} D_{A}/4}[X_{1}^{q}-X_{3}^{q}]+\re^{\beta_{c} D_{A}/4}[X_{2}^{q}-1]=0
\end{eqnarray}
and
\begin{eqnarray}
\label{delta-def21}
M_{B}&=&7\re^{49\beta_{c} D_{B}/4}[Y_{1}^{q}-Y_{5}^{q}]+5\re^{25\beta_{c} D_{B}/4}[Y_{2}^{q}-Y_{6}^{q}]\nonumber \\ 
&+&3\re^{9\beta_{c} D_{B}/4}[Y_{3}^{q}-Y_{7}^{q}]
+ \re^{\beta_{c} D_{B}/4}[Y_{4}^{q}-1]=0.
\end{eqnarray}
The conditions~(\ref{delta-def20}) and~(\ref{delta-def21}) are satisfied when
$X_{1}=X_{3}$ and $X_{2}=1$, while $Y_{1}= Y_{5}$, $Y_{2}= Y_{6}$, $Y_{3}= Y_{7}$ and $Y_{4}= 1$.

Explicitly, we have
\begin{eqnarray}
\label{delta-def22}
X_{1}=X_{3}&=\big[\re^{49\beta_{c} D_{B}/4}\cosh (21\beta_{c} J/4)Y_{1}^{p}+\re^{25\beta_{c} D_{B}/4} \cosh (15\beta_{c} J/4)Y_{2}^{p}\nonumber\\ \nonumber
&+\re^{9\beta_{c} D_{B}/4 }\cosh (9\beta_{c} J/4)Y_{3}^{p}
\re^{\beta_{c} D_{B}/4}\cosh (3\beta_{c} J/4)\big] \\ \nonumber
&\times\big[\re^{49\beta D_{B}/4}\cosh (7\beta_{c} J/4)Y_{1}^{p}
+ \re^{25\beta_{c} D_{B}/4} \cosh (5\beta_{c} J/4)Y_{2}^{p}\\ 
&+\re^{9\beta_{c} D_{B}/4 }\cosh (3\beta_{c} J/4)Y_{3}^{p}+\re^{\beta_{c} D_{B}/4}\cosh (\beta_{c} J/4)\big]^{-1},
\end{eqnarray}
\begin{eqnarray}
\label{delta-def23}
Y_{1}=Y_{5}&=&\big[\re^{9\beta_{c} D_{A}/4 }\cosh (21\beta_{c} J/4)X_{1}^{p}+\re^{\beta_{c} D_{A}/4 }\cosh (7\beta_{c} J/4)\big] \nonumber\\  &\times& \big[\re^{9\beta_{c} D_{A}/4 }\cosh (3\beta_{c} J/4)X_{1}^{p}+\re^{\beta_{c} D_{A}/4 }\cosh (\beta_{c} J/4)\big]^{-1},
\end{eqnarray}
\begin{eqnarray}
\label{delta-def24}
Y_{2}=Y_{6}&=&\big[\re^{9\beta_{c} D_{A}/4 }\cosh (15\beta_{c} J/4)X_{1}^{p}+\re^{\beta_{c} D_{A}/4 }\cosh (5\beta_{c} J/4)\big]\nonumber\\ 
&\times& \big[\re^{9\beta_{c} D_{A}/4 }\cosh (3\beta_{c} J/4)X_{1}^{p}+\re^{\beta_{c} D_{A}/4 }\cosh (\beta_{c} J/4)\big]^{-1},
\end{eqnarray}
and
\begin{eqnarray}
\label{delta-def25}
Y_{3}=Y_{7}&=&\big[\re^{9\beta_{c} D_{A}/4 }\cosh (9\beta_{c} J/4)X_{1}^{p}+\re^{\beta_{c} D_{A}/4 }\cosh (3\beta_{c} J/4)\big]\nonumber\\ 
&\times& \big[\re^{9\beta_{c} D_{A}/4 }\cosh (3\beta_{c} J/4)X_{1}^{p}+\re^{\beta_{c} D_{A}/4 }\cosh (\beta_{c} J/4)\big]^{-1}.
\end{eqnarray}
In order to elucidate the compensation behavior in the model, we introduce the total magnetization expressed as follows:
\begin{eqnarray}
\label{delta-def26}
M_{T}=\frac{1}{2}\left( M_{A}+M_{B}\right).
\end{eqnarray}
For this purpose, we recall that the compensation temperature, denoted as $T_{\rm comp}$, is the temperature at which the total magnetization of the system vanishes before the transition. It is also the temperature at which the absolute values of the sublattice magnetizations are equal, i.e., at $T_{\rm comp}$
\begin{eqnarray}
\label{delta-def27}
M_{T}(T_{\rm comp})=0 \quad {\rm and} \quad |M_{A}(T_{\rm comp})|=|M_{B}(T_{\rm comp})|\neq 0.
\end{eqnarray}

\section{Discussions of the numerical findings}
\label{sec-III}
\subsection{Ground states phase diagram}
In order to determine the existence domains of the various ground states of our system, we have constructed the phase diagram in the $ (D_{A}/q|J|,D_{B}/q|J|)$ plane in the absence of magnetic field ($h=0$). These ground states correspond to the states of minimum energy and are obtained by comparing the values of the internal energy $H_0$ per site for different spin configurations. This energy $H_0$ can be expressed as follows:
\begin{eqnarray}
\label{delta-def28}
H_0= s_{i}\sigma_{j} - \frac{1}{q|J|}\big(D_{A}s_{i}^{2}+D_{B}\sigma_{j}^{2}\big).
\end{eqnarray}
Taking into account the ferrimagnetic interaction ($J <$ 0) and the possible values for the spins of type-${s}$ and type-${\sigma}$, we find eight spin configurations; namely $\big(\frac{3}{2}$, $-\frac{7}{2}\big)$, $\big(\frac{3}{2}$, $-\frac{5}{2}\big)$, $\big(\frac{3}{2}$, $-\frac{3}{2}\big)$, $\big(\frac{3}{2}$, $-\frac{1}{2}\big)$, $\big(\frac{1}{2}$, $-\frac{7}{2}\big)$, $\big(\frac{1}{2}$, $-\frac{5}{2}\big)$, $\big(\frac{1}{2}$, $-\frac{3}{2}\big)$, and $\big(\frac{1}{2}$, $-\frac{1}{2}\big)$.

\begin{figure}[htb]
\begin{center}
	{\includegraphics[scale=0.5]{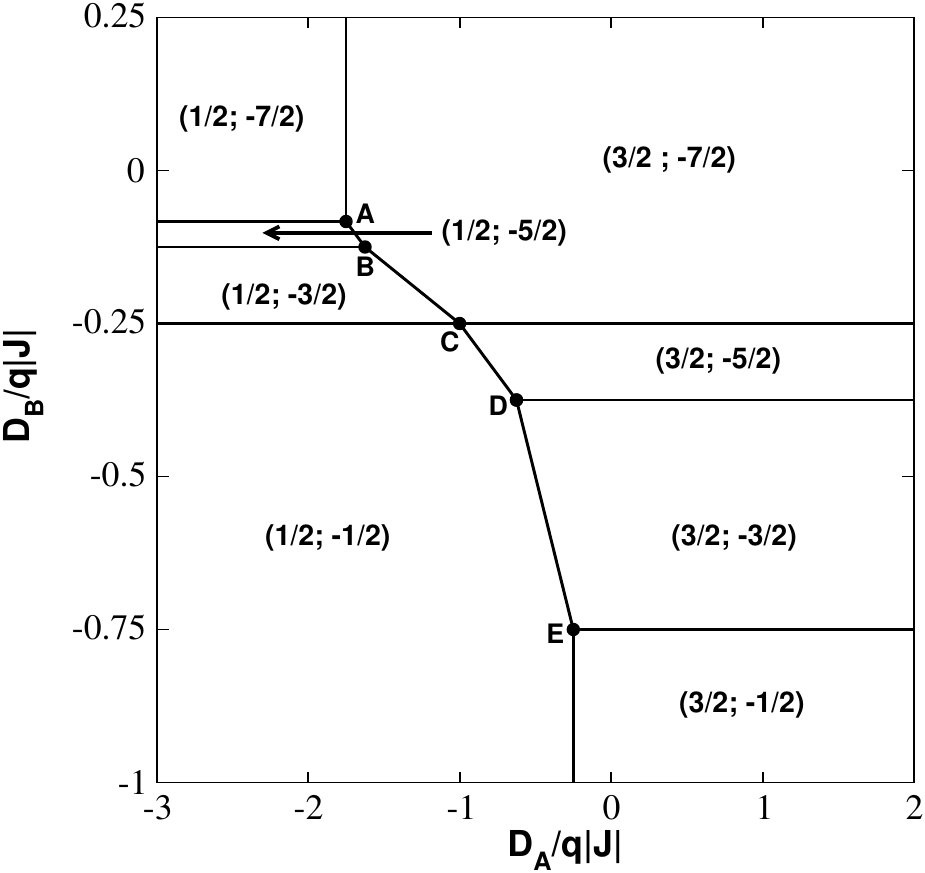}}
\caption{Ground states phase diagram of the model in the $ (D_{A}/q|J|,D_{B}/q|J|)$ plane. Here, $q$ denotes the coordination number.}
 \label{fig-smp1}
\end{center}
\end{figure}

The comparison of energies for these different spin configurations has led to the ground state phase diagram constructed in the $ (D_{A}/q|J|, D_{B}/q|J|) $ plane for all values of the coordination number $q$ in  figure~\ref{fig-smp1}. In this figure, one can observe the existence of multiple phases lines as well as the multicritical points; namely A $\big(-\frac{7}{4}$, $-\frac{1}{12}\big)$, B $\big(-\frac{13}{8}$, $-\frac{1}{8}\big)$, C $\big(-1$, $-\frac{7}{4}\big)$, D $\big(-\frac{5}{8}$, $-\frac{3}{8}\big)$ and E $\big(-\frac{1}{4}$, $-\frac{3}{4}\big)$, where at least three phases coexist. This initial result is in excellent agreement with the one reported in reference~\cite{ra34} where the ground state phase diagram of the same model is constructed in the same plane but for $q = 4$.

\subsection{Thermal and compensation behaviors}
\begin{figure}[h!]
	\begin{center}
		{\includegraphics[scale=0.60]{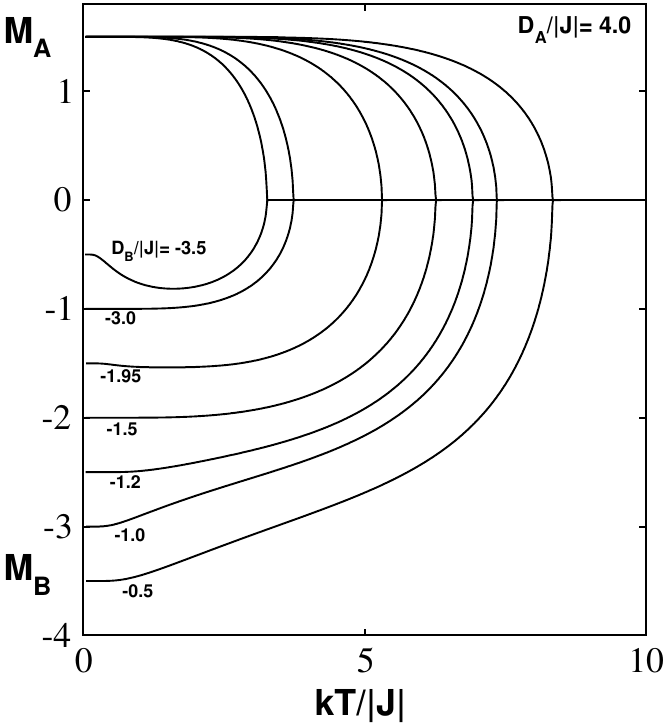}}
		\caption{Thermal behaviors of the sublattice magnetizations $M_{A}$ and $M_{B}$ for $D_{A}/|J|$ = 4.0, $q=4$ and given values of $D_{B}/|J|$ as indicated on the curves.}
		\label{fig-smp2}
	\end{center}
\end{figure}

The previous section devoted to the construction of the ground states phase diagram allowed us to identify the various ground states exhibited by our system, as well as their range of existence. These will be utilized in the current section to verify the saturation values of the magnetizations $M_{A}$ and $M_{B}$ of the two sublattices of the system. The magnetization curves plotted here, obtained through numerical resolution of equations~(\ref{delta-def20})--(\ref{delta-def21}), have been the subject to the analysis for various values of $D_{A}/|J|$ and $D_{B}/|J|$ in the absence of the magnetic field ($h/|J|=0$). The initial thermal variations of the magnetizations analyzed are illustrated in figure~\ref{fig-smp2} for $D_{A}/|J| = 4.0$ and selected values of $D_{B}/|J|$ as indicated in the figure. As can be observed, $M_{A}$
starts from its saturation value of $\frac{3}{2}$ for all values of $D_{B}/|J|$, while $M_{B}$ exhibits at $T =0$ seven saturation values: $-\frac{1}{2}$, $-1$, $-\frac{3}{2}$, $-2$, $-\frac{5}{2}$, $-3$, and $-\frac{7}{2}$ respectively for $D_{B}/|J|= -3.5, -3.0, -1.95, -1.5, -1.2, -1.0$, and $-0.50$. Note that for each value of $D_{B}/|J|$, the magnetization value $M_{A}$ decreases, while $M_{B}$ increases with temperature and both eventually vanish
continuously at a common critical temperature $T_{c}$, which increases with the increase of $D_{B}/|J|$.

\begin{figure}[h!]
	\begin{center}
		{\includegraphics[scale=0.62]{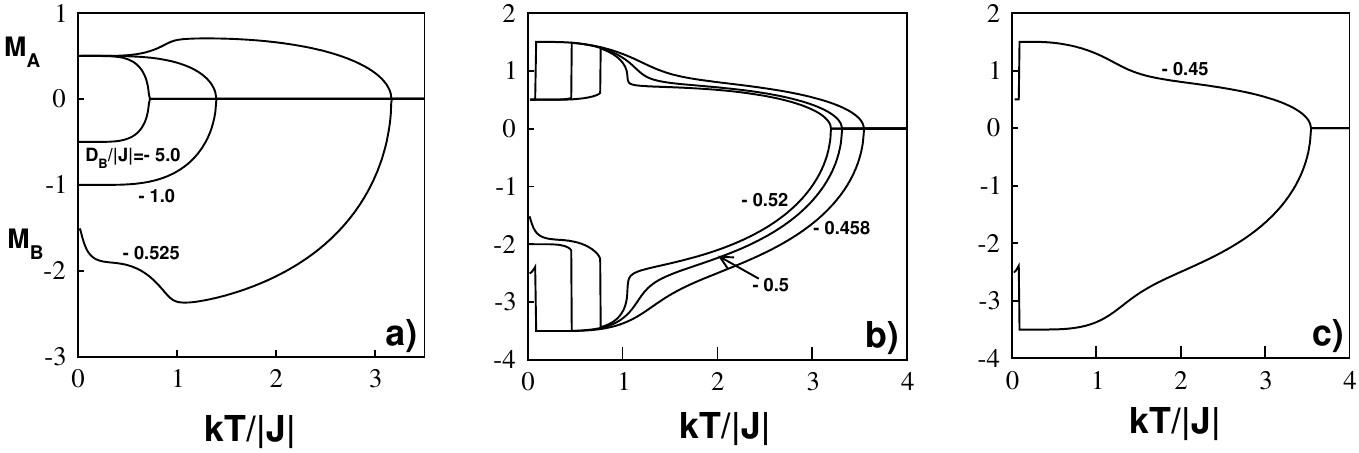}}
		\caption{Thermal behaviors of the sublattice magnetizations $M_{A}$ and $M_{B}$ for $D_{A}/|J| =-6$, $q=4$ and selected values of $D_{B}/|J|$ as indicated on the curves. Here, the model exhibits second- and first-order phase transition temperatures.} \label{fig-smp3}
	\end{center}
	
\end{figure}

When $D_{A}/|J| = -6.0$ (see figure~\ref{fig-smp3}), we can clearly observe in panel a, i.e., for $D_{B}/|J|<  -0.52$, the magnetizations curves exhibit thermal behaviors similar to those observed previously, with the only difference that the magnetization $M_{A}$ starts from its saturation value of $\frac{1}{2}$ for all values of $D_{B}/|J|$. For $-0.52 \leqslant D_{B}/|J|<  -0.45$ ( figure~\ref{fig-smp3}b), the system also exhibits first-order phase transition. Indeed, within this range of $D_{B}/|J|$, there has been a first-order transition between the phases: $\big(\frac{1}{2}, -\frac{5}{2}\big)$ and $\big(\frac{3}{2}, -\frac{7}{2}\big)$; $\big(\frac{1}{2}, -2\big)$ and $\big(\frac{3}{2}, -\frac{7}{2}\big)$; $\big(\frac{1}{2}, -\frac{3}{2}\big)$ and $\big(\frac{3}{2}, -\frac{7}{2}\big)$, where the first-order temperature denoted as $T_{t}$ decreases as $D_{B}/|J|$ increases. For values of $D_{B}/|J|\geqslant  -0.45$ (figure~\ref{fig-smp3}c), the model only shows second-order phase transition where only the phase $\big(\frac{3}{2}, -\frac{7}{2}\big)$ persists.

\begin{figure}[h!]
	\begin{center}
		{\includegraphics[scale=0.60]{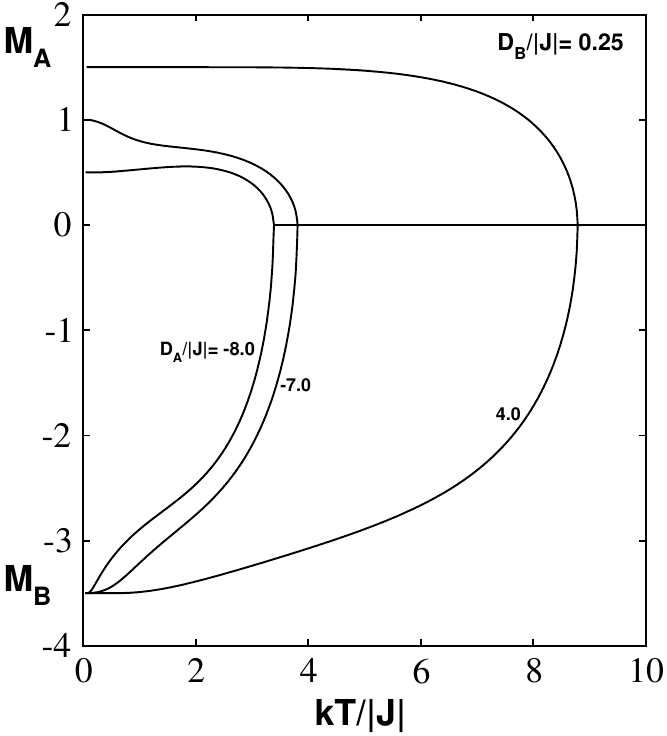}}
		\caption{Thermal behaviors of the sublattice magnetizations $M_{A}$ and $M_{B}$ for $D_{B}/|J| =0.25$, $q=4$ and selected values of $D_{A}/|J|$ as indicated on the curves.} \label{fig-smp4}
	\end{center}
\end{figure}

In figure~\ref{fig-smp4}, where the thermal behaviors of the magnetizations are still being analyzed, for $D_{B}/|J| = 0.25$, the magnetization of sublattice $A$ exhibits three saturation values: $\frac{1}{2}$, $1$, and $\frac{3}{2}$, respectively, for $D_{A}/|J| = -8.0, -7.0$, and 4.0, while $M_{B} = -\frac{7}{2}$ at $T= 0$ for these same parameters. These various results found in figures~\ref{fig-smp2},~\ref{fig-smp3}  and~\ref{fig-smp4} are in perfect agreement with the ground states phase diagram and reveal the key critical behaviors of the model for specific values of the system
parameters. These behaviors are qualitatively similar to those reported in reference~\cite{ra35}, where another model of mixed half-integer spins is studied.

\begin{figure}[!]
\begin{center}
	{\includegraphics[scale=0.55]{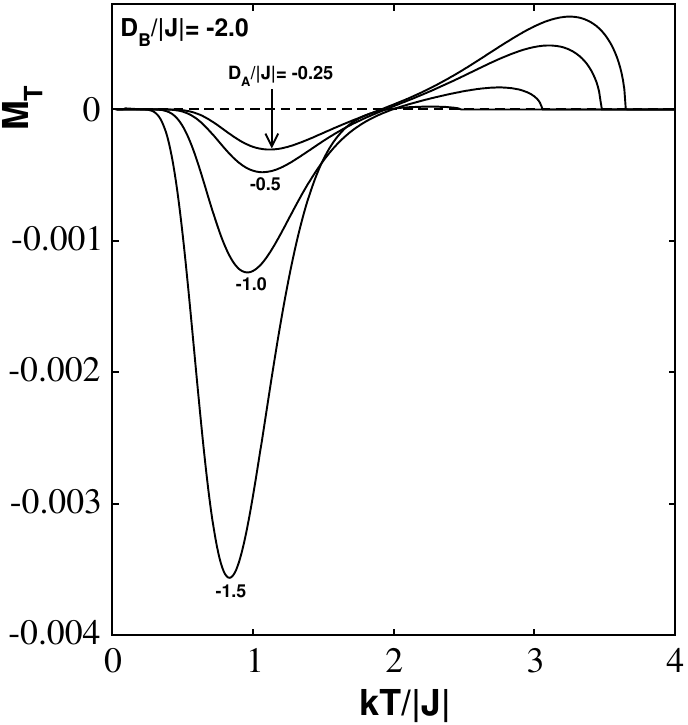}}
\caption{Thermal behaviors of the total magnetization $M_{T}$ for $D_{B}/|J| =-2$, $q=4$ and selected values of $D_{A}/|J|$ as indicated on the curves. The model shows only one compensation temperature for specific values of the model parameters.} \label{fig-smp5}
\end{center}
\end{figure}

In order to clarify the existence of compensation behavior in the system, we have examined the thermal variations of the total magnetization $M_{T}$. As shown in figure~\ref{fig-smp5} calculated for $D_{B}/|J| = -2.0$ and $-1.5 < D_{A}/|J| < 0$, the thermal behaviors of the total magnetization reveal the presence of the compensation phenomenon in the system, since the curves of the total magnetization nullify once before the phase transition.

Furthermore, all these curves almost pass through the same compensation point. We can thus conclude that the compensation behavior is not affected by the increase in the value of $D_{A}/|J|$, while the critical temperature $T_{c}$ increases with the increase of $D_{A}/|J|$. This latter result is in complete agreement with the one found in reference~\cite{ra34}, with the difference that the critical temperatures are lower than those reported in the cited reference, as expected.

\begin{figure}[h!]
	\begin{center}
		{\includegraphics[scale=0.55]{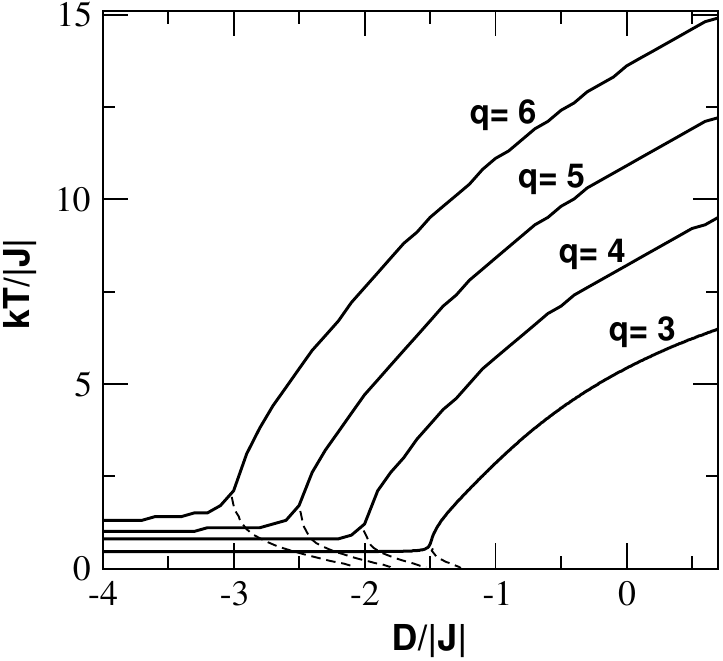}}
		\caption{Finite temperature phase diagrams in the ($D/|J|, kT_{c}/|J|$) plane for $q=3, 4, 5$ and $6$.  The solid and dashed lines refer to the second- and first-order phase transition lines, respectively.} \label{fig-smp6}
	\end{center}
\end{figure}

\subsection{Phase diagrams }
In this section, we present the phase diagrams obtained from the analysis of the thermal variations of the magnetizations $ M_{A} $, $ M_{B} $, and $ M_{T} $. Thus, we present the phase diagrams in the  ($D/|J|, kT/|J|$) plane, in the ($D_{A}/|J|, kT/|J|$) plane as well as in the  ($D_{B}/|J|, kT/|J|$) plane. In these diagrams of interest, the solid and dashed lines represent the second-order and first-order
transition lines, respectively. In addition, the compensation lines are indicated by dotted-dashed lines.

The first phase diagram, as shown in figure~\ref{fig-smp6}, is displayed in the ($D/|J|, kT/|J|$) plane for the given values of the coordination number $q$.  The second-order transition lines separating the ordered ferrimagnetic phase from the disordered paramagnetic phase start at low temperatures for all
$q$ and are nearly horizontal for large negative values of $D/|J|$ until near $D/|J| = -\frac{q}{2}$.  After this specific value of $D/|J|$, these transition lines increase with the increase of $D/|J|$, and as this latter tends towards infinity, they once again become nearly horizontal. As for the first-order transition lines, they start around $D/|J| = -\frac{q}{2}$ and at high temperatures near the corresponding second-order
transition line, then gradually decrease until they disappear as $D/|J|$ increases. It should also be noted that each of these transition lines, for each value of $q$, separates the ferrimagnetic phase $\big(\frac{1}{2}, -\frac{1}{2}\big)$ from the ferrimagnetic phases $\big(\frac{1}{2}, -\frac{3}{2}\big)$ or $\big(\frac{1}{2}, -2\big)$ or $\big(\frac{1}{2}, -\frac{5}{2}\big)$ or $\big(\frac{1}{2}, -3\big)$ or even $\big(\frac{1}{2}, -\frac{7}{2}\big)$ depending on whether
$D/|J|$ tends towards large positive values. It is important to mention that the found results in this figure show a certain resemblance with those of references~\cite{ra23,ra35}.

\begin{figure}[h!]
\begin{center}
	{\includegraphics[scale=0.6]{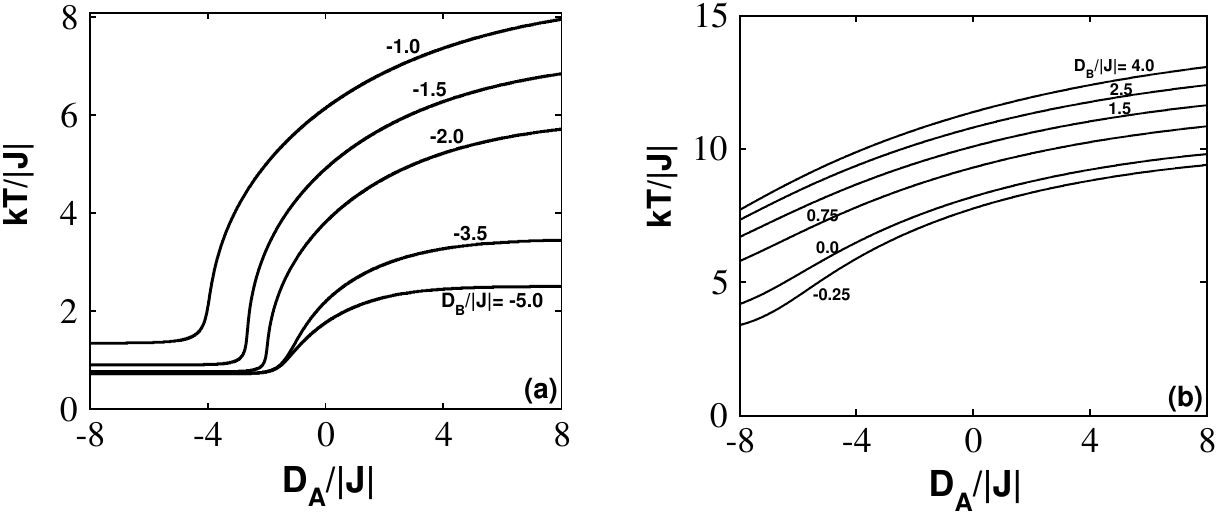}}
\caption{Finite temperature phase diagrams in the ($D_A/|J|, kT_{c}/|J|$) plane for $q=4$ and given values of $D_B/|J|$ as indicated on the transition lines.} \label{fig-smp7}
\end{center}
\end{figure}

The next phase diagram is calculated, as shown in figure~\ref{fig-smp7}(a--b), in the ($ D_{A}/|J|,kT/|J|$) plane with varying values of $D_{B}/|J|$. For  $D_{B}/|J|< -0.25$ (figure~\ref{fig-smp7}a), the critical lines exhibit similar behaviors to those of the previous figure. On the other hand, when $D_{B}/|J|\geqslant -0.25$, i.e., in figure~\ref{fig-smp7}b, these critical lines start at high temperature and increase exponentially as $D_{A}/|J|$ approaches infinity. It is also worth noting that the critical temperature increases with a
 simultaneous increase of $D_{A}/|J|$ and $D_{B}/|J|$.

\begin{figure}[h!]
	\begin{center}
		{\includegraphics[scale=0.53]{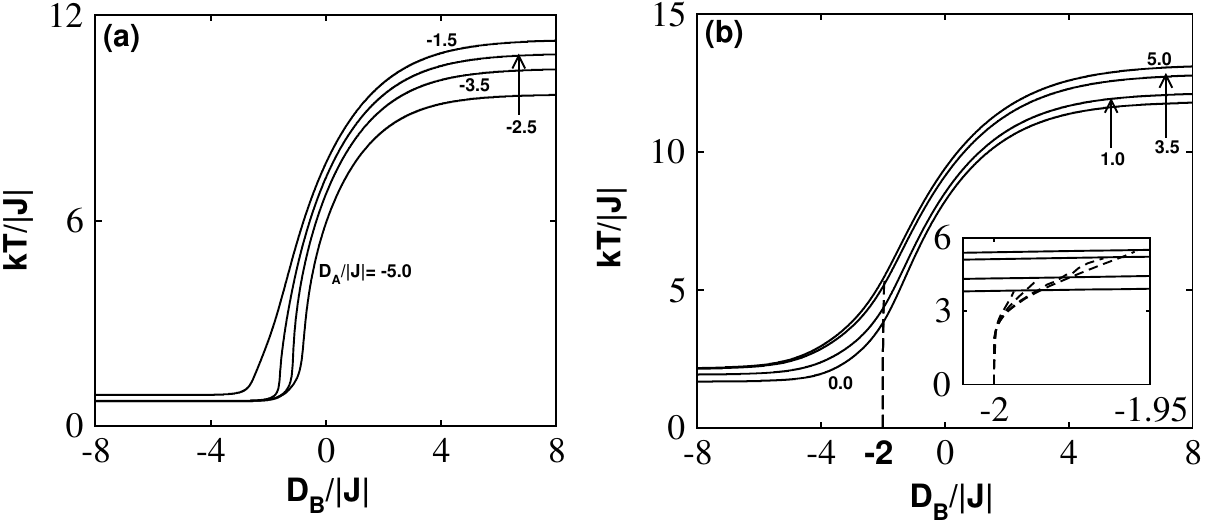}}
		\caption{Finite temperature phase diagrams in the ($D_B/|J|, kT_{c}/|J|$) and ($D_B/|J|, kT_{\rm comp}/|J|$) planes combined for $q=4$ and given values of $D_A/|J|$ as indicated on the transition lines. Here, the solid and dotted-dashed lines refer to the second-order phase transition and compensation lines, respectively.} \label{fig-smp8}
	\end{center}
\end{figure}

The last phase diagram depicted in figure~\ref{fig-smp8} of this manuscript is constructed in the ($D_{B}/|J|, kT/|J|$) plane with variation of the $D_{A}/|J|$ values in each case. As shown in figure~\ref{fig-smp8}a, for $D_{A}/|J|< 0$, the critical lines exhibit similarity to those in figure~\ref{fig-smp7}a, with the only difference that the critical temperatures are higher here. As observed in the inset of figure~\ref{fig-smp8}b, for all
values of $D_{A}/|J|$ ($D_{A}/|J| > 0$), all compensation lines originate from the specific value $D_{B}/|J| = -2.0$ and at zero temperature, and then increase with the increase of $D_{B}/|J|$ to terminate near the corresponding critical line. However, the critical behavior remains the same as in the previous cases.

\subsection{Hysteresis properties}
In this section, we have presented the various results obtained concerning the hysteresis behavior of the system. For this purpose, we have analyzed the influence of parameters such as temperature, anisotropies, and the ferrimagnetic coupling parameter on this hysteresis behavior.

\begin{figure}[h!]
	\begin{center}
		{\includegraphics[scale=0.53]{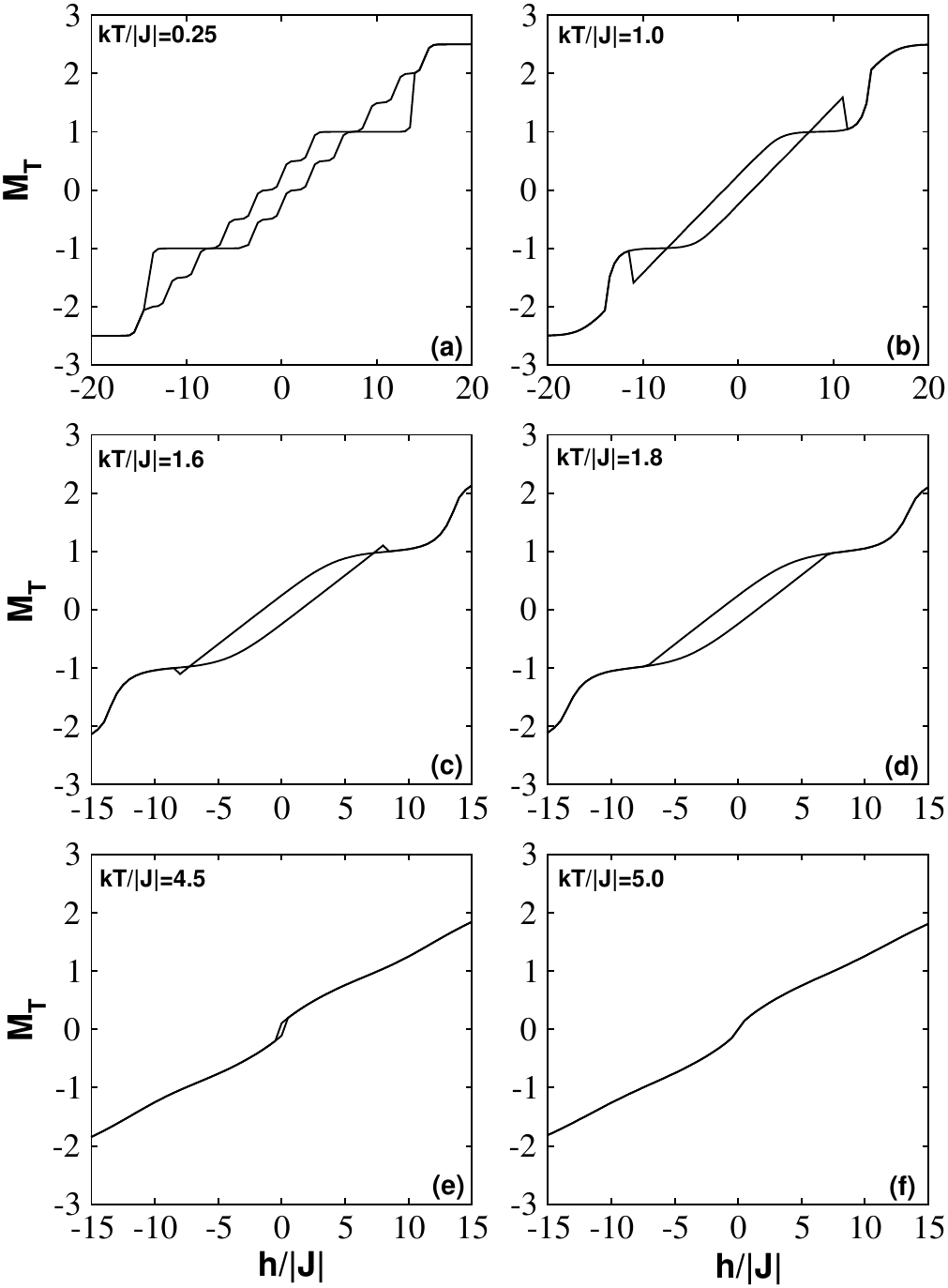}}
		\caption{Hysteresis properties of the model for $D_{A}/|J|=0$, $D_{B}/|J|=-1.5$ and  given values of the temperature as shown in the figure.} \label{fig-smp9}
	\end{center}
\end{figure}

Initially, we investigated the influence of temperature on the hysteresis behavior in the system with the fixed value of a crystal field acting on the sites of one sublattice, while the second one was considered to be zero. Figure~\ref{fig-smp9} obtained for $D_{A}/|J| = 0$ and $D_{B}/|J| = -1.5$ reveals the existence of a multi-hysteresis behavior in the system. Indeed, the number of hysteresis loops changes from three to one before disappearing as the temperature increases. As depicted in this figure, when $kT/|J|< 1.8$ (figure~\ref{fig-smp9}a--c), the system exhibits three hysteresis loops, including a central loop connected to two lateral loops that gradually vanish as the temperature rises. In figure~\ref{fig-smp9}(d--e), i.e., for $1.8 \leqslant kT/|J| \leqslant 4.5$, only one hysteresis loop is observed, which already disappears when $kT/|J| = 5.0$ (figure~\ref{fig-smp9}f), this latter temperature being higher than the critical temperature ($kT_{c}/|J| =4.9099$) at which the system transforms into its paramagnetic phase for these same parameter values.

\begin{figure}[h!]
\begin{center}
	{\includegraphics[scale=0.6]{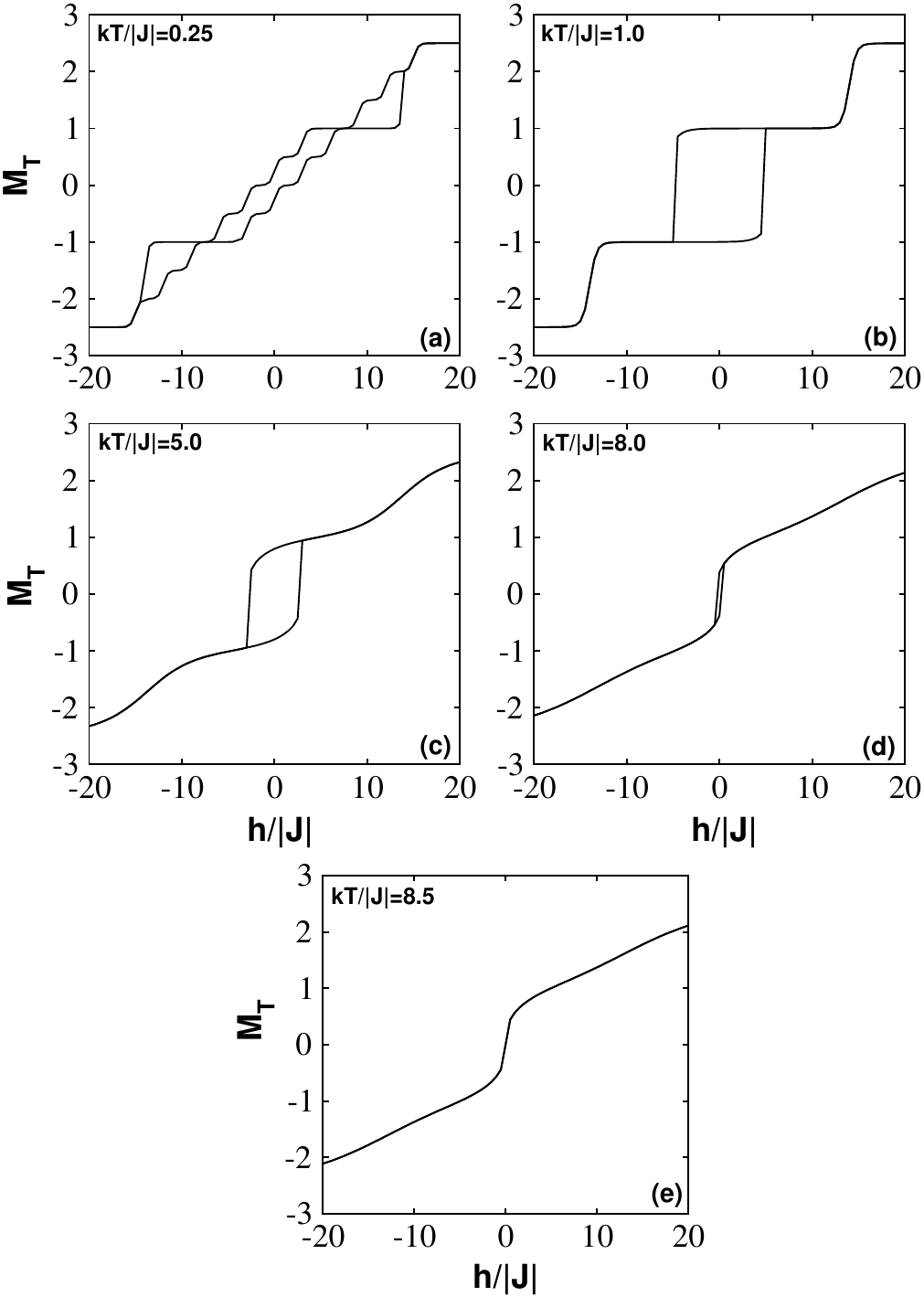}}
\caption{Hysteresis properties of the model for $D_{A}/|J|=1.0$, $D_{B}/|J|=0$ and given values of the temperature as shown in the figure.} \label{fig-smp10}
\end{center}
\end{figure}

In figure~\ref{fig-smp10}, depicted for $D_{A}/|J| = 1.0$ and $D_{B}/|J| = 0$, the observed hysteresis behaviors are similar to those analyzed in the previous figure, with the only difference being that they appear in different temperature ranges. Indeed, the multi-hysteresis behavior emerges for $kT/|J|< 1.0$ (figure~\ref{fig-smp10}a). The sole loop, with decreasing size and width, is observed for $1.0 \leqslant kT/|J| \leqslant 8.0$. At $kT/|J| = 8.5$, the hysteresis behavior disappears for the same reason mentioned in figure~\ref{fig-smp9}f.

\begin{figure}[h!]
\begin{center}
	{\includegraphics[scale=0.6]{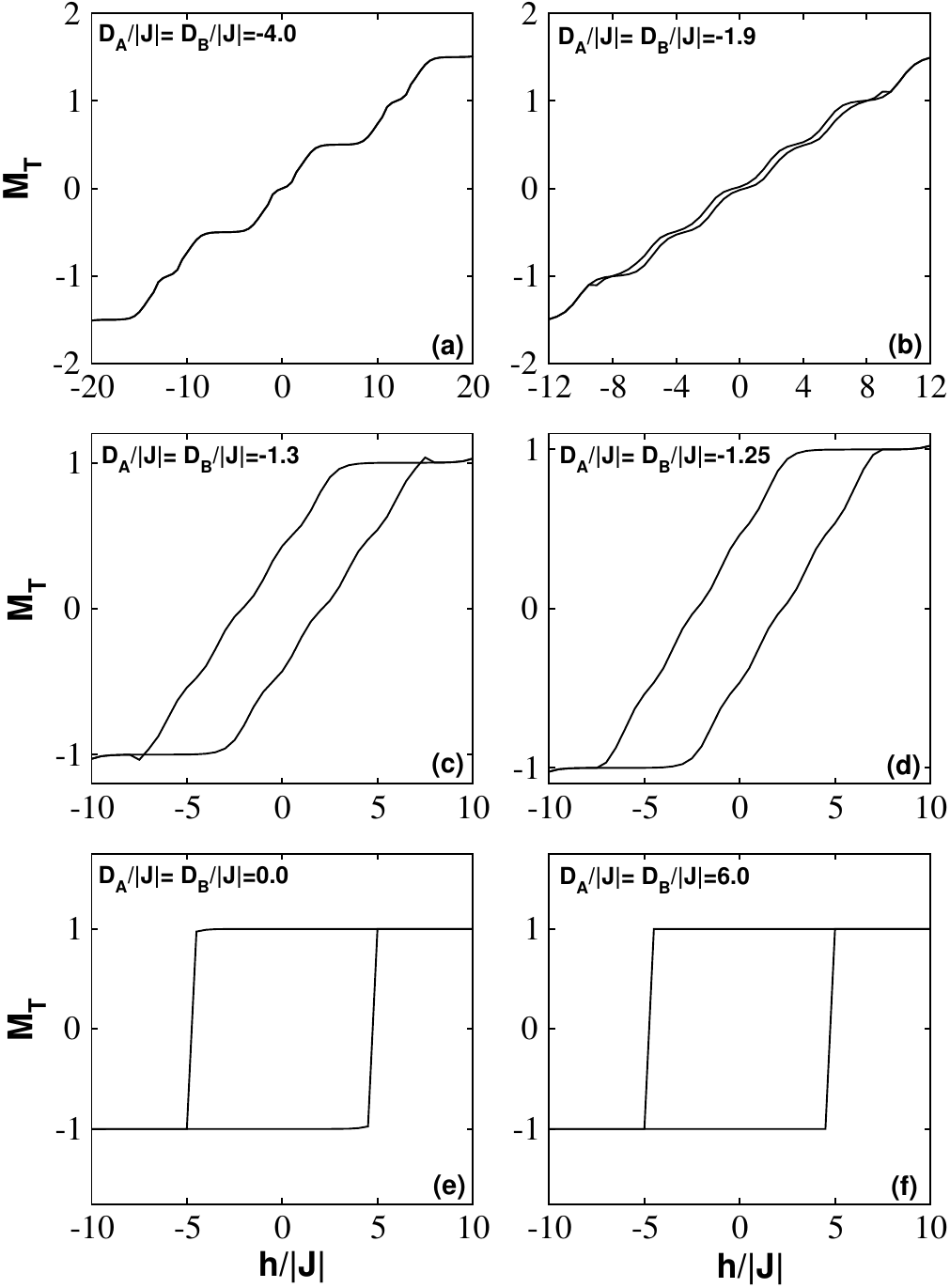}}
\caption{Hysteresis properties of the model for $kT/|J|=0.5$ and selected values of $D_{A}/|J|=D_{B}/|J|$ as shown in the figure.} \label{fig-smp11}
\end{center}
\end{figure}

Figure~\ref{fig-smp11} exhibits hysteresis behaviors for the fixed temperature value ($kT/|J| = 0.5$) and the given values of the crystal field chosen uniformly for the sublattices, i.e., $D_{A}/|J| = D_{B}/|J| = D/|J|$.  As can be observed in this figure, the uniform crystal field has an impact on the hysteresis behavior in our system. Indeed, when $D/|J|< -1.9$, no hysteresis loop appeared, while for
$-1.9 \leqslant D/|J|< -1.25$ and $D/|J| \geqslant  -1.25$, respectively, three and one loop are observed with widths and sizes that remain constant when $D/|J| \geqslant  0$. Thus, the coercive field and the remanent magnetization are not affected by the increase in the positive uniform crystal field.

\begin{figure}[h!]
\begin{center}
	{\includegraphics[scale=0.6]{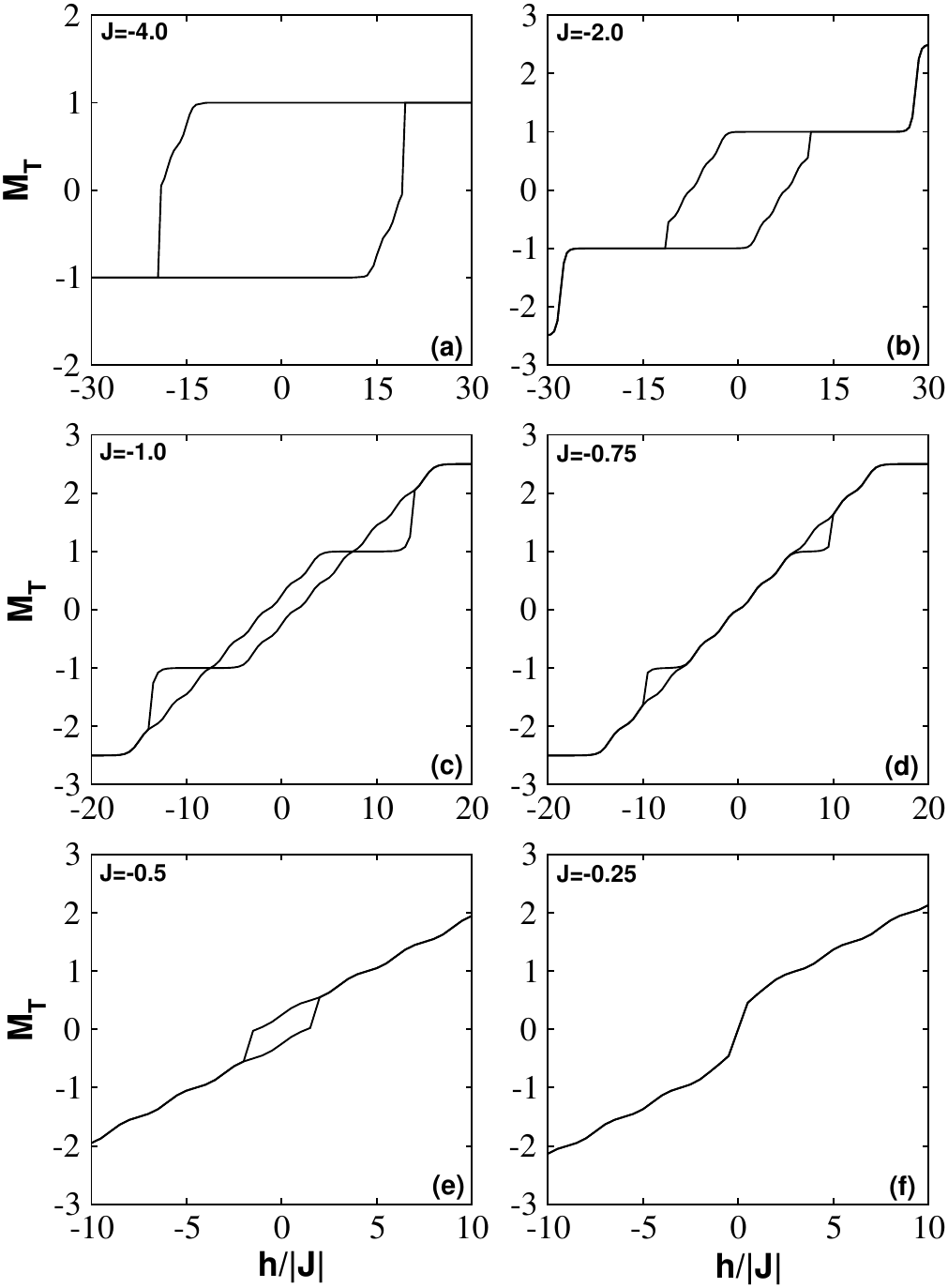}}
\caption{Impacts of the ferrimagnetic coupling $J$ on the hysteresis loops of the  model for $ D_{A}/|J|= 0$, $D_{B}/|J|=-1.5$ and $kT/|J|=0.5$.} \label{fig-smp12}
\end{center}
\end{figure}

Finally, we also investigated the influence of the ferrimagnetic interaction $J$ on the same behavior in the system. As depicted in figure~\ref{fig-smp12}, which illustrates the total magnetization as a function of the external magnetic field $h$, the multi-hysteresis behavior is indeed exhibited in the system for other parameters, i.e., for $ D_{A}/|J| = 0$, $ D_{B}/|J| = -1.5$ and $kT/|J| = 0.5$. For $J < -1.0$ (figure~\ref{fig-smp12}a--b), only a single hysteresis cycle is observed. The case where three hysteresis loops appear, i.e., for $-1.0 \leqslant J < -0.75$ (figure~\ref{fig-smp12}c), is extensively discussed in the previous figures. As the value of the ferrimagnetic interaction increases, the width and size of the central loop observed in figure~\ref{fig-smp12}c shrink, and it disappears, leaving only the two lateral loops, as evident for $J = -0.75$ (figure~\ref{fig-smp12}d). These two lateral loops, under the influence of the increased magnetic interaction, vanish, and a single central cycle reemerges when $-0.5 \leqslant J < -0.25$, only to disappear again for values of $J \geqslant -0.25$. This final result has allowed us to clearly observe that the interaction coupling has an impact on the size and width of the hysteresis loops.Thus, there is a dependence between the interaction coupling and the coercitive field $h_{c}$ on one hand, and the remanent magnetization $M_{R}$ on the other hand; the coercitive field $h_{c}$ and the remanent magnetization $M_{R}$ being two important characteristics to comprehend the magnetic behavior of materials.

\begin{figure}[h!]
\begin{center}
	{\includegraphics[scale=0.6]{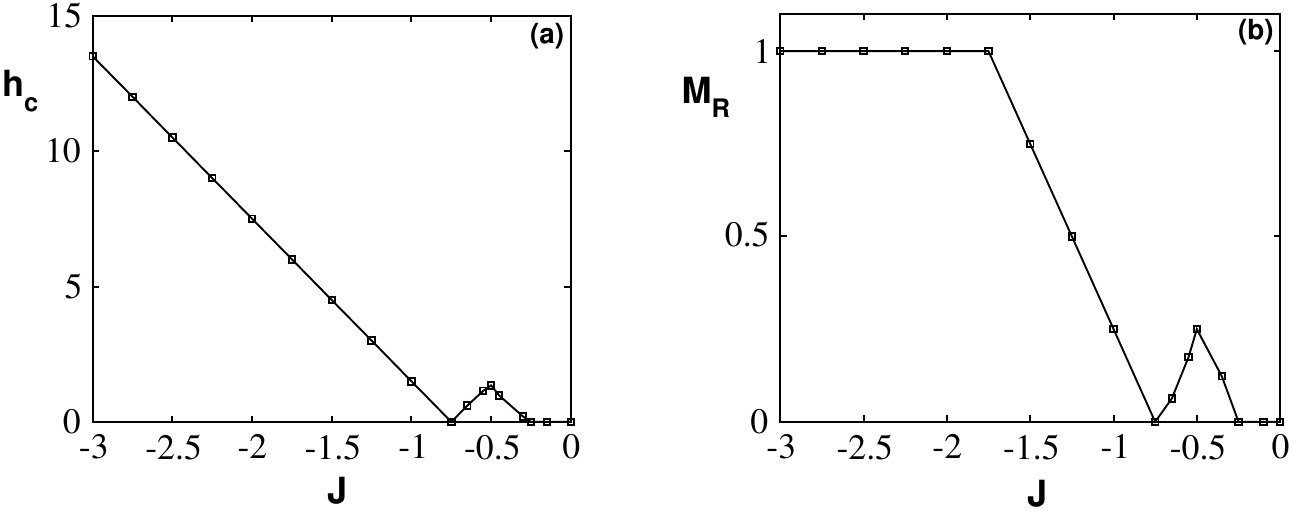}}
\caption{Impacts of the ferrimagnetic coupling $J$ on the coercitive field and remanent magnetization of the  model for  $ D_{A}/|J|= 0$, $D_{B}/|J|=-1.5$ and $kT/|J|=0.5$.} \label{fig-smp13}
\end{center}
\end{figure}

In figure~\ref{fig-smp13}, we depict the dependence between the ferrimagnetic interaction coupling $J$ and the coercive field $h_{c}$ and remanent magnetization $(M_{R})$. In figure~\ref{fig-smp13}a, when $J \leqslant -0.75$, the coercive field gradually decreases and becomes zero at $J = -0.75$. For $-0.75 < J \leqslant -0.5$, the coercive field changes direction, i.e., it increases progressively and reaches a maximum value ($h_{\rm cmax} = 1.35$) at $J = -0.50$. Beyond this last value of $J$, the coercive field gradually decreases to become zero definitively starting from $J = -0.25$. From figure~\ref{fig-smp13}b, it is evident that $M_{R} = 1.0$ indicating that the ferrimagnetic interaction coupling does not affect the remanence when $J \leqslant -1.75$. Beyond this value of $J$, the remanence exhibits a behavior similar to that of the coercive field observed in figure~\ref{fig-smp13}a. These results illustrated in figure~\ref{fig-smp13} are valuable for making appropriate parameter choices depending on whether one intends to use this type of material as a soft material (low coercivity) which finds applications in the design of transformers, inductors, electromagnet cores, and other electromagnetic components, or as a hard material (high coercivity) used in various applications, including speakers, electric motors, hard disks, etc. This last result aligns perfectly with the ones reported in reference~\cite{ra20}.

\section{Conclusion}
\label{sec-IV}

In this paper, we have investigated the critical, compensation, and hysteresis behaviors  of a  mixed spin-3/2 and 7/2 ferrimagnetic Ising system on the Bethe lattice using the method based on exact recursion relations. Through the thermal variations of magnetizations and phase diagrams in multiple planes, we have demonstrated that the studied system exhibits first- and second-order transition as well as compensation behaviors. Furthermore, we have discovered that the system, under the influence of an external magnetic field, exhibits one, two, or three hysteresis loops, the size and width of which depend on the system parameters. It is thus evident that the system displays a multi-hysteresis properties. Comparing our results with those existing in the literature especially to those reported in references~\cite{ra34} where the same model is investigated for $q=4$ corresponding to the square lattice by means of mean field approximation, our results are not only in perfect agreement with their results but also more interesting. 

{It is worth noting that the compensation and multi-hysteresis properties exhibited by this study offer opportunities for designing materials and devices with specific magnetic characteristics, suitable for a variety of technological applications. A thorough understanding of these properties enables us to optimize the performance of magnetic materials and expand the possibilities in fields such as data storage and magnetic sensors. As our  future work, we plan to investigate in detail the effects of the interaction coupling between second neighboring atoms on the magnetic properties revealed by the present study by using not only the recursion relations technique but also the Monte Carlo simulations technique which is known as one of the most reliable methods for studying complex systems.}

\lastpage


\ukrainianpart

\title{Дослідження критичної, компенсаційної та гістерезисної поведінки у феромагнітній моделі Блюма--Капеля із напівцілим спіном-(3/2, 7/2): точні рекурсивні співвідношення}
\author{М. Каке\refaddr{label1}, С.~І.~В. Хонтінфінде\refaddr{label2, label3,label5}, М. Каріму\refaddr{label1,label4,label5},  Р.~Хуену\refaddr{label1}, Е.~Албайрак\refaddr{label6}, Р.~А.~А.~Єссуфу\refaddr{label1,label3}, A. Кпадону\refaddr{label1,label7}
}
\addresses{
		\addr{label1} Інститут фізико-математичних наук, Дангбо, Бенін
	\addr{label2} Вища національна технологічна школа математичного моделювання Абомей, Бенін
	\addr{label3} Фізичний факультет університету Абомей-Калаві, Бенін
	\addr{label4} Лабораторія технічних наук і прикладної математики
	\addr{label5} Вища національна школа енергетики та технічних наук, Абомей, Бенін
	\addr{label6}  Фізичний факультет університету Ерджіес, 38039, Кайсері, Туреччина
	\addr{label7} Лабораторія прикладної фізики, Абомей, Бенін
}

\makeukrtitle

\begin{abstract}
	\tolerance=3000%
Точні рекурсивні співвідношення застосовано для дослідження феримагнітної системи Блюма--Капела Ізінга зі змішаним напівцілим спіном-(3/2, 7/2) на ґратці Бете. Для виявлення різних можливих основних станів моделі, в площині $({D_{A}}/{q|J|}, {D_{B}}/{q|J|})$ отримано фазові діаграми основного стану. Використовуючи температурну залежність параметрів порядку, побудовано досить цікаві фазові діаграми у площинах $(D_{A}/|J|, kT/|J|)$, $(D_{B}/|J|,\ , kT/|J|)$ та $(D/|J|,\, kT/|J|)$, де $D=D_A=D_B$. Виявлено, що для певних значень параметрів моделі в системі мають місце фазові переходи першого та другого роду, а також існують точки компенсації. При певних умовах на значення зовнішнього магнітного поля модель також має багато-гістерезисну поведінку з наявністю одинарного, подвійного та потрійного циклів.  Досліджено вплив феримагнітного зв’язку $J$ на залишкову намагніченість і коерцитивні поля для певних значень фізичних параметрів системи. Отримані числові результати якісно узгоджуються з даними, відомими з наукової літератури.
	\keywords рекурсивні співвідношення, феримагнітна система Блюма--Капеля, основний стан, ґратка Бете, петлі гістерезису
	
\end{abstract}

\end{document}